\documentclass[twocolumn,aps,prb,reprint,showpacs,amsmath,amssymb]{revtex4-1}

\usepackage{epsfig}
\usepackage{graphicx}
\usepackage{dcolumn}
\usepackage{bm}
\usepackage[colorlinks=true,dvipdfm]{hyperref}
\usepackage{multirow}
\usepackage{CJK}
\usepackage{float}

\begin{document}
\begin{CJK*}{GBK}{}
\title{Hybridized Kibble-Zurek scaling in the driven critical dynamics across an overlapping critical region}
\author{Liang-Jun Zhai$^{1}$} \email{zhailiangjun@jsut.edu.cn}
\author{Huai-Yu Wang$^{2}$}
\author{Shuai Yin$^{3}$} \email{zsuyinshuai@163.com}
\affiliation{$^{1}$The school of mathematics and physics, Jiangsu University of Technology, Changzhou 213001, China}
\affiliation{$^{2}$Department of Physics, Tsinghua University, Beijing 100084, China}
\affiliation{$^{3}$Institute for Advanced Study, Tsinghua University, Beijing, 100084, China}
\date{\today}

\begin{abstract}
The conventional Kibble-Zurek scaling describes the scaling behavior in the driven dynamics across a single critical region. In this paper, we study the driven dynamics across an overlapping critical region, in which a critical region (Region-A) is overlaid by another critical region (Region-B). We develop a hybridized Kibble-Zurek scaling (HKZS) to characterize the scaling behavior in the driven process. According to the HKZS, the driven dynamics in the overlapping region can be described by the critical theories for both Region-A and Region-B simultaneously. This results in a constraint on the scaling function in the overlapping critical region. We take the quantum Ising chain in an imaginary longitudinal-field as an example. In this model, the critical region of the Yang-Lee edge singularity and the critical region of the ferromagnetic-paramagnetic phase transition point overlap with each other. We numerically confirm the HKZS by simulating the driven dynamics in this overlapping region. The HKZSs in other models are also discussed.

\end{abstract}

\maketitle

\section{\label{intro}Introduction}
Understanding exotic phenomena in the critical region is one of the most fascinating arena in condensed matter physics~\cite{Cardy,Sachdev,Sachdev2}. Near the critical point, long-wavelength modes dominate the macroscopic behavior of the system. As a result, critical systems always show universal scaling properties and the dependence of macroscopic quantities on relevant variables is usually characterized by power-law functions~\cite{Cardy,Sachdev,Sondhi}. These universal power-law functions also manifest themselves in the nonequilibrium dynamics near the critical point~\cite{Tauber}. For instance, for the relaxation dynamics, the theory of the short-time critical dynamics has demonstrated that the relaxation process exhibits a universal critical initial slip characterized by an additional dynamic exponent~\cite{Janssen,Janssen2,Zheng,Yin,Zhang,Shu}; while for the driven dynamics across a critical point, the Kibble-Zurek mechanism predicts that topological defects emerge after the quench and the number of the topological defects is a power-law function of the driving rate~\cite{Kibble,Zurek,Dz,Pol}. Moreover, the Kibble-Zurek scaling (KZS) shows that only the equilibrium critical exponents are needed to characterise the driven dynamics scaling~\cite{Kibble,Zurek,Dz,Pol}. Recently, lots of efforts have been made to examine and generalize the Kibble-Zurek mechanism in various systems~\cite{qkz1,qkz2,qkz3,qkz4,BDamski,qkz5,qkz6,qkz7,qkz8,qkz9,Chandran,Francuz,qkz10,Zhong1,Zhong2,NDAntunes,Bermudez,Bermudez1,Shimizu}. For instance, the Kibble-Zurek mechanism has been verified in trapped-ion systems and Bose-Einstein condensates~\cite{Ulm,Pyka,Navon,Clark}. Theoretically, the full scaling form of the KZS has been developed in both classical and quantum phase transitions~\cite{Zhong1,Zhong2,qkz3,qkz6,qkz10}. Furthermore, this full scaling form has been employed to numerically detect the critical properties in lots of systems~\cite{Zhong1,Zhong2,CWLiu1,CWLiu2,qkz10,Hu}.

In addition to the quantum and classical phase transitions, the driven critical dynamics has also been studied in the dissipative phase transition of the Yang-Lee edge singularity (YLES)~\cite{Yang,Lee,Kortman,Fisher,YinYL}. It has been shown than the KZS is still applicable in the YLES, although the Kibble-Zurek mechanism for the generation of the topological defects breaks down. A feature of the KZS therein is that the driven process crosses an overlapping region constituted by the critical regions of the $(0+1)$-dimensional ($(0+1)$D) YLES and the $(1+1)$D YLES~\cite{YinYL}. In this overlapping region, it was found that the driven dynamics can be described by the KZS with two sets of critical exponents determined by both the $(0+1)$D and the $(1+1)$D YLESs.

Besides the YLES in the quantum Ising chain, overlapping critical regions also appear in lots of other systems. The most familiar one is the $2$D quantum Ising model at finite temperatures~\cite{Cardy,Sachdev}. In this model, the $(2+1)$D quantum Ising critical region and the $2$D classical Ising critical region overlap with each other~\cite{Cardy,Sachdev}. Another example is the $2$D Dirac system coupling with a spin model at finite temperatures~\cite{Stephanov,Hesselmann}. In this model, the $(2+1)$D chiral Ising critical region and the $2$D classical Ising critical region overlap. Since overlapping critical regions are very common phenomena in condensed matter systems, a general scaling theory on the driven dynamics across the overlapping critical region is called for.

In this paper, we develop a hybridized Kibble-Zurek scaling (HKZS) theory to describe the driven dynamics across the overlapping critical region, which is constituted by critical region A (Region-A) and critical region B (Region-B). The HKZS asserts that (1) the driven dynamics can be described by the critical theories in both Region-A and Region-B; (2) the scaling forms in the overlapping region should satisfy a constraint, which includes the information in both Region-A and Region-B. In the main text, we take the $1$D quantum Ising model in an imaginary longitudinal-field as an example. This model is the same as the one studied in Ref.~\cite{YinYL}. However, in this paper, we study the overlapping region contributed by the critical regions of the $(1+1)$D ferromagnetic-paramagnetic phase transition (FPPT) and the $(0+1)$D YLES. After numerically confirming the HKZS, we discuss the applications of the HKZS in other systems.

The rest of the paper is organized as follows. In Sec.~\ref{sec:YLES}, we first give an introduction about the phase diagram of the quantum Ising model in an imaginary longitudinal field, followed by a discussion on its static critical scaling behaviors. Then, we illustrate the theory of the HKZS in Sec.~\ref{sec:HKZS}. Numerical verifications are presented in Sec.~\ref{sec:Numerialresult}. Then the HKZS in other models is discussed in Sec.~\ref{discu} and a summary is given in Sec~\ref{sum}.

\section{\label{sec:YLES}Model and its static scaling behavior}
\subsection{\label{sec:semodel}The quantum Ising model in an imaginary longitudinal-field}
The Hamiltonian of the quantum Ising chain in an imaginary longitudinal-field reads
\begin{eqnarray}
  \mathcal{H} &=& -\sum_{n=1}^{L}{\sigma_n^z\sigma_{n+1}^z}-\lambda \sum_{n=1}^{L}{\sigma_n^x}-ih\sum_{n=1}^{L}\sigma_{n}^z,
  \label{model}
\end{eqnarray}
where $\sigma_n^z$ and $\sigma_n^x$ are the Pauli matrices at $n$ site in the $z$ and $x$ direction, respectively, $\lambda$ is the transverse field, $h$ is imaginary-part of the longitudinal field, and $L$ is the lattice size. For $L=\infty$, model~(\ref{model}) exhibits an FPPT at $(g,h)=(g_c,h_c)\equiv(0,0)$, where $g\equiv(\lambda-\lambda_c)$ and $\lambda_c=1$~\cite{Cardy,Sachdev}. Besides this usual $(1+1)$D phase transition, for any $g>0$ (referred to as $g_{\rm YL}^L$), there are also critical points for the $(0+1)$D YLES at $(g_{\rm YL}^L,h_{\rm YL}^L)$ for different $L$~\cite{Uzelac,Gehlen}. Note that in contrast to the FPPT, which must occur in the thermodynamic limit, the $(0+1)$D YLES occurs at finite size and the location of its critical point depends on the lattice size $L$~\cite{Uzelac,Gehlen}, as indicated in the superscript.

To be consonant with the definition of the order parameter in the $1$D classical YLES, the order parameters for the $(0+1)$D quantum YLES should be defined as~\cite{Fisher,Uzelac,YinYL}
\begin{eqnarray}
  M_R &=& \mathrm{Re}[\langle\Psi^*|\hat{M}|\Psi\rangle/\langle\Psi^*|\Psi\rangle], \nonumber \\
  M_I &=& \mathrm{Im}[\langle\Psi^*|\hat{M}|\Psi\rangle/\langle\Psi^*|\Psi\rangle],
  \label{Eq:orderp}
\end{eqnarray}
in which $\hat{M}=\sum_{n}^{L}\sigma_n^z/L$, $|\Psi\rangle$ is the wave function, and $M_R$ ($M_I$) is the real (imaginary) part of $\langle\hat{M}\rangle$.

The order parameters defined in Eq.~(\ref{Eq:orderp}) accommodate the phase transition information in both the YLES and the FPPT. In the YLES, for a fixed $g_{\rm YL}^L$ ($g_{\rm YL}^L>0$), the system is in one phase with $M_R=0$ and $M_I\neq0$ when $h<h_{\rm YL}^L$, since the real part of model (\ref{model}) is dominated and the energy spectra are real; when $h>h_{\rm YL}^L$ the system is in the other phase with $M_R\neq0$ and $M_I=0$, since the dissipative part of model (\ref{model}) dominates and the energy spectra are conjugate pairs~\cite{Gehlen}. In the FPPT, $|\Psi\rangle$ can be chosen to be a real function (up to an arbitrary global phase). So $M_R$ returns to the usual definition of the FPPT order parameter, which is nonzero in the ferromagnetic phase and is zero in the paramagnetic phase, and $M_I$ is zero in both phases~\cite{Sachdev}.

\subsection{\label{sec:staticsca}The static scaling behavior}
In this section, we show the static scaling properties of model~(\ref{model}). For the $(0+1)$D YLES, at a fixed $g_{\rm YL}^L$, $M_R$ and $M_I$ satisfy~\cite{Kortman,Fisher,Uzelac,Gehlen},
\begin{eqnarray}
  M_R(h-h_{\rm YL}^L) &\propto& (h-h_{\rm YL}^L)^{\frac{1}{\delta_0}}, \nonumber \\
  M_I(h-h_{\rm YL}^L) &\propto& (h-h_{\rm YL}^L)^{\frac{1}{\delta_0}},
  \label{Eq:orderps1}
\end{eqnarray}
in which $\delta_0=-2$~\cite{Kortman,Fisher} (For the sake of clarity, we list all the relevant exponents in Table \ref{tabexp}). Since $\delta_0$ is negative, $M_R$ and $M_I$ diverge at $(g_{\rm YL}^L,h_{\rm YL}^L)$.

\begin{figure}
  \centerline{\epsfig{file=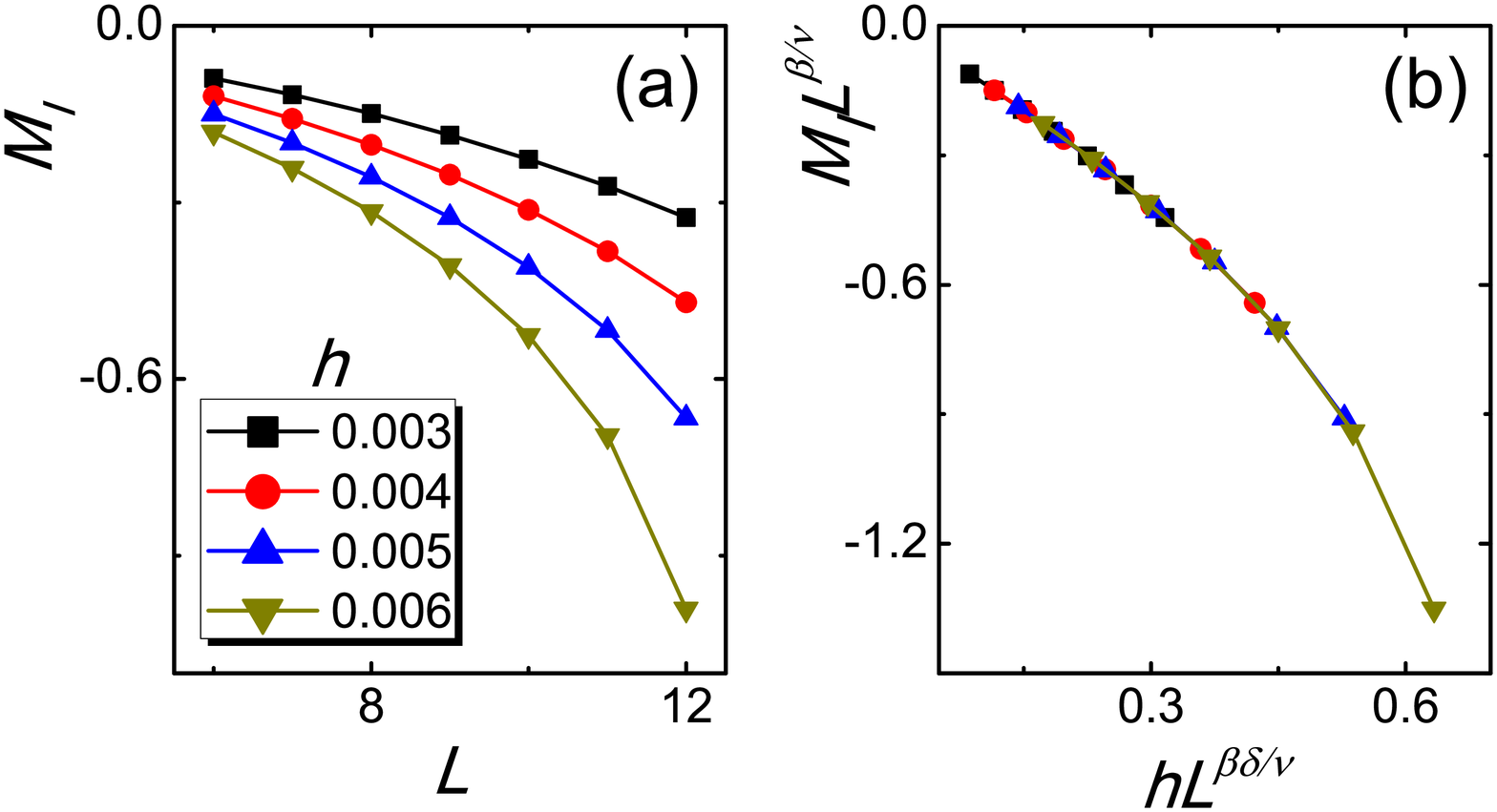,width=0.9\columnwidth}}
  \caption{\label{staticops} (Color online) For fixed $gL^{1/\nu}=0.01$, the curves of $M_I$ versus $L$ for various $h$ in (a) match with each other in (b) after rescaling according to Eq.~(\ref{Eq:orderps3}).}
\end{figure}

For the $(1+1)$D FPPT, the imaginary longitudinal-field has the same dimension as the real longitudinal-field. As a result, similar to the real longitudinal-field case, the scaling forms of the order parameter for a system with size $L$ read~\cite{Sachdev,Sondhi}
\begin{eqnarray}
  M_R(g,h,L) &=& L^{-\frac{\beta}{\nu}}f_1(gL^{\frac{1}{\nu}},hL^{\frac{\beta \delta}{\nu}}), \nonumber \\
  M_I(g,h,L) &=& L^{-\frac{\beta}{\nu}}f_2(gL^{\frac{1}{\nu}},hL^{\frac{\beta \delta}{\nu}}),
  \label{Eq:orderps3}
\end{eqnarray}
in which $\beta$, $\delta$, and $\nu$ are usual critical exponents for the $2$D classical Ising universality class (See Table \ref{tabexp}), and $f_i$ is a scaling function (similar definitions will always be implied). To the best of our knowledge, Eq.~(\ref{Eq:orderps3}) has not been confirmed for an imaginary longitudinal-field. We numerically confirm Eq.~(\ref{Eq:orderps3}) in Fig. \ref{staticops}.
\begin{table}
  \centering
  \caption{Critical exponents for the $(0+1)$D YLES and the $(1+1)$D FPPT, respectively.}
    \begin{tabular}{c p{0.6cm} p{0.6cm} p{0.6cm} p{0.6cm} p{0.6cm} p{0.6cm} p{0.6cm} p{0.6cm} p{0.6cm}}
    \hline
    \hline
    \multicolumn{1}{c|}{YLES} &$\nu_0$ &$\beta_0$ & $\delta_0$ & $z_0$ & $r_0$ &$\frac{\beta_0}{\nu_0}$ &$\frac{\beta_0\delta_0}{\nu_0}$ &$\frac{\beta_0}{\nu_0 r_0}$ &$\frac{\beta_0 \delta_0}{\nu_0 r_0}$ \\
    \hline
    \multicolumn{1}{c|}{} &-1   &1 & -2  & 1 & 3 &-1 &$2$ & $-\frac{1}{3}$ & $\frac{2}{3}$ \\
    \hline
    \hline
    \multicolumn{1}{c|}{FPPT} &$\nu$ &$\beta$ & $\delta$  & $z$ & $r$ &$\frac{\beta}{\nu}$ &$\frac{\beta\delta}{\nu}$ & $\frac{\beta}{\nu r}$ &$\frac{\beta\delta}{\nu r}$  \\
    \hline
    \multicolumn{1}{c|}{} &1  &$\frac{1}{8}$ & 15  & 1 & $\frac{23}{8}$ &$\frac{1}{8}$ &$\frac{15}{8}$ &$\frac{1}{23}$ &$\frac{15}{23}$\\
    \hline
    \end{tabular}%
  \label{tabexp}%
\end{table}%

\begin{figure}
  \centering
  \includegraphics[width=2.5 in]{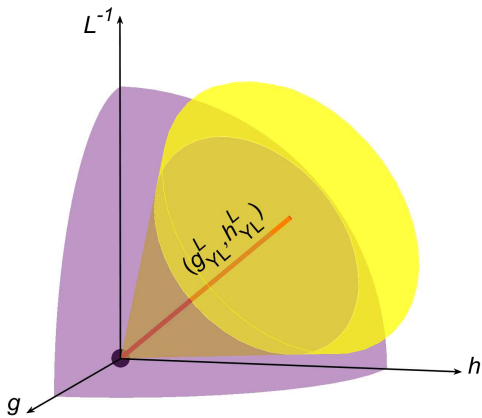}\\
  \caption{Critical regions near the FPPT critical point (origin). Critical points of the $(0+1)$D YLES $(g_{\rm YL}^L,h_{\rm YL}^L)$ link up into a curve (Red-boldface curve), which terminates at the FPPT critical point. The critical region of the $(0+1)$D YLES (yellow cone) thrusts into the critical region of the FPPT. We show that the driven critical dynamics in the overlapping region can be described by the HKZS.}\label{Fig:phase}
\end{figure}

\begin{figure}
  \centerline{\epsfig{file=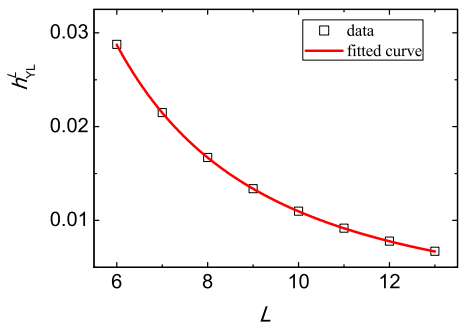,width=0.7\columnwidth}}
  \caption{\label{SizeeffectHz} (Color online) Near the FPPT critical point, the curve of $h_{\rm YL}^L$ versus $L$ for fixed $g_{\rm YL}^L L^{1/\nu}=0.06$. Power-law fitting shows that the curve satisfies $h_{\rm YL}^L\propto L^{-1.885}$, which agrees with Eq.~(\ref{Eq:HYLL}).}
\end{figure}

A remarkable property for model~(\ref{model}) is that for large $L$ and small $g$, the critical region of the $(0+1)$D YLES and the critical region of the $(1+1)$D FPPT overlap unavoidably with each other, as sketched in Fig.~\ref{Fig:phase}. Near the FPPT critical point (the origin in Fig.~\ref{Fig:phase}), the divergence of the order parameter at $(g_{\rm YL}^L,h_{\rm YL}^L)$ gives a constraint on the values of $(g_{\rm YL}^L,h_{\rm YL}^L)$. To see this, by substituting $(g_{\rm YL}^L,h_{\rm YL}^L)$ into Eq.~(\ref{Eq:orderps3}), one finds that $f_{1,2}(g_{\rm YL}^LL^{\frac{1}{\nu}},h_{\rm YL}^LL^{\frac{\beta \delta}{\nu}})=\infty$. This leads to
\begin{equation}
  \label{Eq:HYLL}
  h_{\rm YL}^L=L^{-\frac{\beta\delta}{\nu}}f_3(g_{\rm YL}^L L^{\frac{1}{\nu}}).
\end{equation}
We confirm Eq.~({\ref{Eq:HYLL}}) numerically in Fig.~\ref{SizeeffectHz}. Note that Eq.~({\ref{Eq:HYLL}}) binds $h_{\rm YL}^L$ and $g_{\rm YL}^L$ together with a FPPT critical exponent. This indicates that the critical properties in the overlapping critical region are affected by both the $(0+1)$D YLES and the $(1+1)$D FPPT.

\section{\label{sec:HKZS}Hybridized Kibble-Zurek scaling}
We consider the driven critical dynamics across the overlapping region by changing $h$ as $h=h_0+Rt$. We will show the hybridized effects induced by the cooperation of the $(0+1)$D YLES and the $(1+1)$D FPPT. We will derive the scaling theory of the HKZS and show how the driven dynamics is characterized by the HKZS. In the following, only the scaling behavior of $M_R$ will be considered since $M_I$ satisfies the same scaling theory.

If we focus on the $(0+1)$D YLES, the driven critical dynamics is described by the usual KZS. We choose $h_0$ to be large enough that it is irrelevant when $R$ is small~\cite{YinChen,Silvi}. For a fixed $g_{\rm YL}^L$, the driven dynamics is characterized by $h$ and $R$. Under the external driving, the scaling form of the order parameter reads~\cite{Zhong1,Zhong2,Chandran}
\begin{equation}
M_R(h-h_{\rm YL}^L,R) = R^{\frac{\beta_0}{\nu_0r_0}}f_{4}[(h-h_{\rm YL}^L)R^{-\frac{\beta_0\delta_0}{\nu_0r_0}}],
\label{Eq:yless}
\end{equation}
in which $r_0=z_0+\beta_0\delta_0/\nu_0$. In classical and Hermitian quantum systems, Eq.~(\ref{Eq:yless}) can be explained by a finite-time scaling theory~\cite{Zhong1,Zhong2,qkz10} by analysing the time-scales in different driven stages. However, in the non-Hermitian system of model~(\ref{model}), the definition of the time-scale is ambiguous, since the Hamiltonian has both real and imaginary parts. Nonetheless, it has been shown that Eq.~(\ref{Eq:yless}) is still applicable in the YLES far away from the critical point~\cite{YinChen}.

On the other hand, if we focus on the $(1+1)$D FPPT, in a similar way, we obtain the scaling form by taking the finite-size effects into account~\cite{Zhong1,Zhong2,Huang,YinZhong}
\begin{equation}
  M_{R}(g,h,L,R) = R^{\frac{\beta}{\nu r}}f_{5}(gR^{-\frac{1}{\nu r}},L^{-1}R^{-\frac{1}{r}},hR^{-\frac{\beta\delta}{\nu r}}),
  \label{Eq:fppts}
\end{equation}
in which $r=z+\beta\delta/\nu$. When $L<R^{-1/r}$, the role played by $R$ is negligible and Eq.~(\ref{Eq:fppts}) returns to Eq.~(\ref{Eq:orderps3}).

The HKZS asserts that the driven dynamics in the overlapping critical region can be described by both Eq.~(\ref{Eq:yless}) and Eq.~(\ref{Eq:fppts}). The reason for this assertion is that both $f_3$ and $f_4$ are analytical functions for any finite $R$ and neither of them breaks down in the critical region. The hybridized effects are then embedded in the relation between Eq.~(\ref{Eq:yless}) and Eq.~(\ref{Eq:fppts}).

To uncover the hinge between them, we start from Eq.~(\ref{Eq:fppts}). By noticing that in Eq.~(\ref{Eq:yless}) $g=g_{\rm YL}^L$ and $L$ is fixed, we set in Eq.~(\ref{Eq:fppts}) $g$ to be $g_{\rm YL}^L$ and $L$ to be a constant. So, $gL^{1/\nu}$ is also a constant. By substituting these into Eq.~(\ref{Eq:fppts}), one finds that
\begin{equation}
  M_{R}(g_{\rm YL}^L,h,L,R) = R^{\frac{\beta}{\nu r}}f_{6}(g_{\rm YL}^LL^{\frac{1}{\nu}},L^{-1}R^{-\frac{1}{r}},hR^{-\frac{\beta\delta}{\nu r}}).
  \label{Eq:fppts1}
\end{equation}
Then, according to Eq.~(\ref{Eq:HYLL}), one can replace $g_{\rm YL}^LL^{1/\nu}$ with $h_{\rm YL}^LL^{\beta\delta/\nu r}$ in Eq.~(\ref{Eq:fppts1}). After changing the variable $h$ to be $h-h_{\rm YL}^L$, one obtains
\begin{equation}
  M_{R}(h-h_{\rm YL}^L,L,R) = R^{\frac{\beta}{\nu r}}f_{7}[L^{-1}R^{-\frac{1}{r}},(h-h_{\rm YL}^L)R^{-\frac{\beta\delta}{\nu r}}],
  \label{Eq:fppts2}
\end{equation}
for a fixed $g_{\rm YL}^L L^{1/\nu}$. Comparing Eq.~(\ref{Eq:fppts2}) with Eq.~(\ref{Eq:yless}), one finds that $f_7(A,B)$ satisfies
\begin{equation}
  f_7(A,B)=(A^{-r})^{\frac{\beta_0}{\nu_0 r_0}-\frac{\beta}{\nu r}}f_{8}[B(A^{r})^{\frac{\beta_0 \delta_0}{\nu_0 r_0}-\frac{\beta \delta}{\nu r}}].
  \label{Eq:fppts3}
\end{equation}
Since $f_7$ is identical to $f_5$ for a fixed $g_{\rm YL}^LL^{1/\nu}$, Eq.~(\ref{Eq:fppts3}) shows that $f_{5,7}$ itself should satisfy a scaling form, which includes the exponents of both the $(0+1)$D YLES and the $(1+1)$D FPPT, bridging the gap between Eq.~(\ref{Eq:yless}) and Eq.~(\ref{Eq:fppts3}). Therefore, Eq.~(\ref{Eq:fppts3}) plays a central role in the HKZS.

For a general case, in which the overlapping critical region is assumed to be constructed by the critical Region-A and critical Region-B, the HKZS asserts:

\textit{(1) In the overlapping critical region, the driven dynamics can be described by the critical theories of both Region-A and Region-B simultaneously;}

\textit{(2) The scaling function itself should satisfy a scaling form which includes the critical information in both Region-A and Region-B.}

\section{\label{sec:Numerialresult}Numerical results}
In this section, we numerically solve the Schr\"{o}dinger equation for model (\ref{model}) to verify the HKZS.

\begin{figure}[h]
  \centerline{\epsfig{file=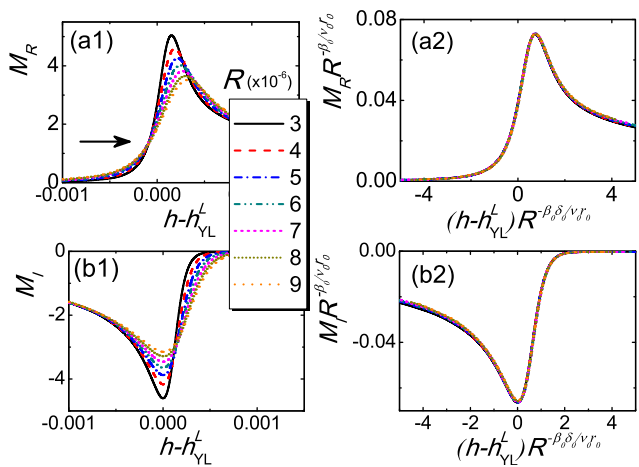,width=1\columnwidth}}
  \caption{Under increasing $h$ with $L=10$, the curves of $M_R$ vs $h-h_{\rm YL}^L$ (a1) match with each other in (a2) when $M_R$ and $h-h_{\rm YL}^L$ are rescaled by the (0+1)D YLES exponents. The corresponding curves for $M_I$ are shown in (b1) and (b2), respectively. $h_0$ is chosen as $h_0=0$. The critical point of the YLES $(g_{\rm YL}^L,h_{\rm YL}^L)=(0.001,0.010558)$. The arrow in (a1) points the direction of changing $h$.}
\label{Fig:hx1001L10}
\end{figure}

First, we examine that, in the overlapping region constructed by the $(0+1)$D YLES and the $(1+1)$D FPPT critical regions, the driven dynamics can be described by the $(0+1)$D YLES, i.e., Eq.~(\ref{Eq:yless}). We choose $L=10$ and $g=0.001$. From Fig.~\ref{staticops}, one finds that this lattice size is large enough that the system has entered the critical region of the $(1+1)$D FPPT. The results are shown in Fig.~\ref{Fig:hx1001L10}. From Fig.~\ref{Fig:hx1001L10}, we find that for fixed $g_{\rm YL}^L$ and $L$, after rescaling $M_R$ and $h-h_{\rm YL}^L$ with $R$ by using the $(0+1)$D YLES exponents, the rescaled curves collapse onto each other, confirming Eq.~(\ref{Eq:yless}). Note that we also calculate the driven dynamics of $M_I$. For $M_I$, the same conclusion is still applicable except that the scaling function is different.

\begin{figure}
  \centerline{\epsfig{file=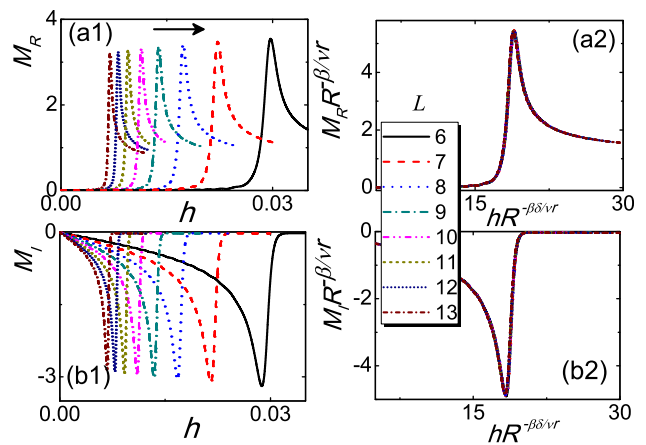,width=1\columnwidth}}
  \caption{Under increasing $h$ for fixed $LR^{1/r}=0.191488$ and $gR^{-1/\nu r}=0.313336$, the curves of $M_R$ vs $h$ (a1) match with each other in (a2) when $M_R$ and $h$ are rescaled by the (1+1)D FPPT exponents. The corresponding curves for $M_I$ are shown in (b1) and (b2), respectively. $h_0$ is chosen as $h_0=0$. The arrow in (a1) points the direction of changing $h$.}
\label{Fig:sizeeffect}
\end{figure}

Second, we examine that, in the overlapping critical region, the driven dynamics can be described by the $(1+1)$D FPPT, i.e., Eq.~(\ref{Eq:fppts}). We calculate the driven dynamics with fixed $LR^{1/r}$ and $gR^{-1/\nu r}$. The curves of $M_R$ versus $h$ are plotted in Fig.~\ref{Fig:sizeeffect} (a1). After rescaling $M_R$ and $h$ with $R$, we find that the rescaled curves match with each other according to Eq.~(\ref{Eq:fppts}) as shown in Fig.~\ref{Fig:sizeeffect} (a2). We also calculate the dynamics of $M_I$. We find that the dynamics of $M_I$ also satisfies Eq.~(\ref{Eq:fppts}) except for a different scaling function.

\begin{figure}
  \centerline{\epsfig{file=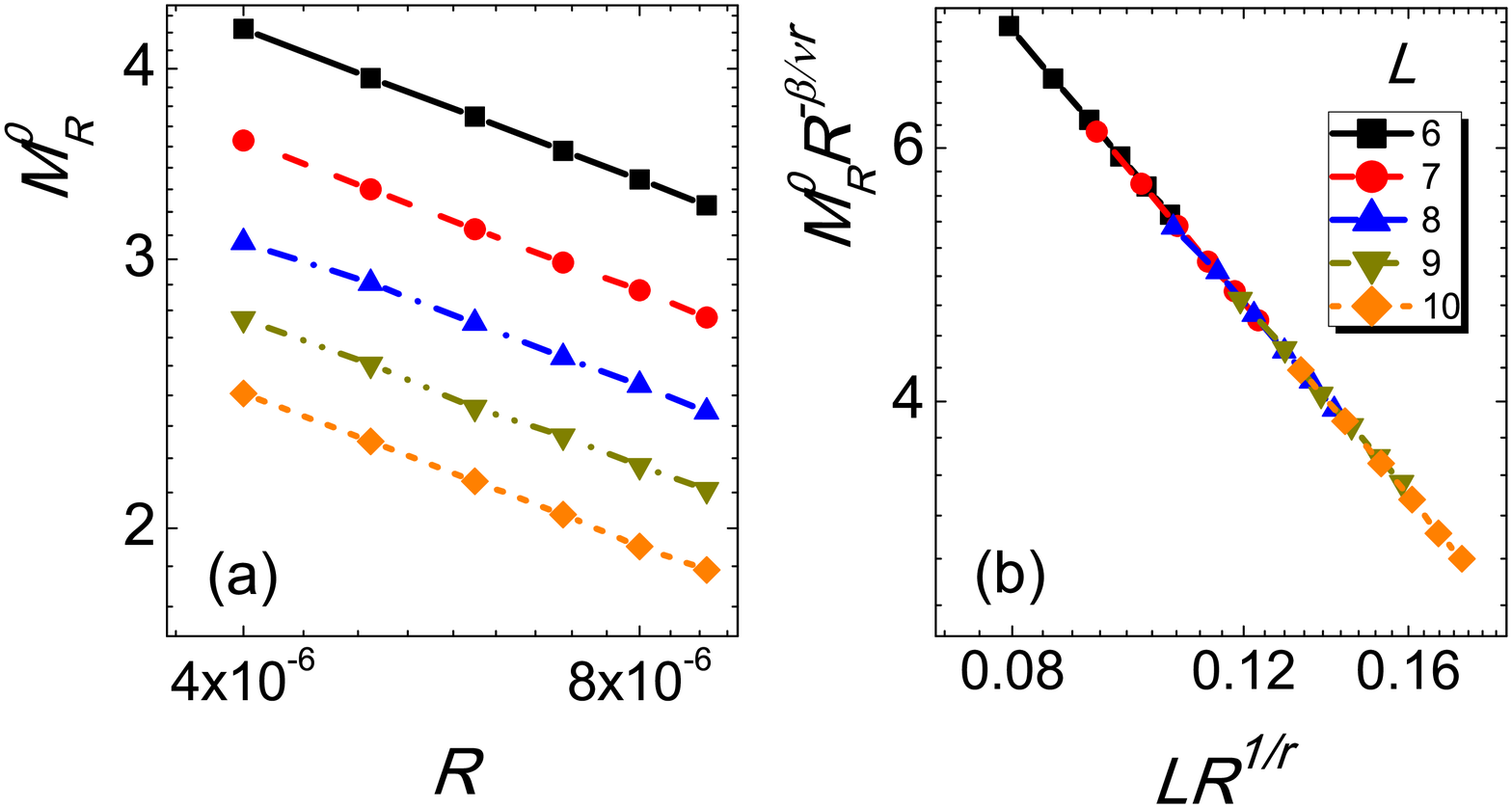,width=1\columnwidth}}
  \caption{Under increasing $h$ for fixed $gL^{1/\nu}=0.06$, (a) curves of $M_R^0$ versus $R$ for different lattice sizes; and (b) the collapse of curves of the rescaled $M_R^0$ versus the rescaled $L$. Double-logrithmic scales are used in both (a) and (b).}
\label{Fig:comrela}
\end{figure}

Third, we check the relation between Eq.~(\ref{Eq:yless}) and Eq.~(\ref{Eq:fppts}) by examining Eq.~(\ref{Eq:fppts3}). To this end, we extract $M_R$ for different $R$ at $h_{\rm YL}^L$ and denote them $M_R^0$. For each $L$, we plot the curve of $M_R^0$ versus $R$ in Fig.~\ref{Fig:comrela} (a). We find that these curves in double-logarithmic scale are almost parallel straight lines with the average slope being $-0.324$, agreeing with the theoretical value of $\beta_0/\nu_0 r_0=-1/3$ (See Table \ref{tabexp}). Then we rescale $M_R^0$ as $M_R^0R^{-\beta/\nu r}$ and plot the results as a function of $LR^{1/r}$ in Fig.~\ref{Fig:comrela} (b). We find these curves collapse onto each other according to Eq.~(\ref{Eq:fppts2}). By plotting the rescaled curve in the double-logarithmic scale, we find that it is almost a straight line, whose slope is about $-1.069$. This exponent is close to $r(\beta_0/\nu_0 r_0-\beta/\nu r)\simeq -1.083$, confirming Eq.~(\ref{Eq:fppts3}).

\section{\label{discu}Discussion}
\subsection{\label{discu1}Two-dimensional quantum Ising model at low temperatures}
Here, we discuss the HKZS in the driven dynamics in the $2$D quantum Ising model at low temperatures. The Hamiltonian is~\cite{Cardy,Sachdev}
\begin{eqnarray}
  \mathcal{H}_2 &=& -\sum_{\langle n,m\rangle}{\sigma_n^z\sigma_{m}^z}-\lambda \sum_{n}{\sigma_n^x}-h\sum_{n}\sigma_{n}^z,
  \label{model2}
\end{eqnarray}
where $\langle\rangle$ denotes the nearest interaction. As we mentioned in Sec.~\ref{intro}, the quantum critical region of the $(2+1)$D FPPT is overlaid by the classical critical region of the $2$D FPPT. We again parameterize the distance to the quantum critical poing $\lambda_c$ as $g\equiv \lambda-\lambda_c$. In the overlapping critical region, the classical phase transition point $T_c$ depends on $g$ as~\cite{Cardy,Sachdev}
\begin{equation}
T_c(g)=|g|^{\nu_q z_q},
\label{Eq:tcg}
\end{equation}
which has been verified numerically in Ref.~\cite{Hesselmann}.

The dynamics of this system is described by the Lindblad equation~\cite{Attal,YinZhong,YinChen2},
\begin{eqnarray}
\begin{split}
\frac{\partial \rho}{\partial
t}=&-i[\mathcal H_2,\rho]-c\sum_{m\neq l}W_{l\rightarrow m}(V_{l\rightarrow m}^\dagger V_{l\rightarrow m} \rho\\&+\rho V_{l\rightarrow m}^\dagger V_{l\rightarrow m}-2V_{l\rightarrow m} \rho V_{l\rightarrow m}^\dagger), \label{Lind}
\end{split}
\end{eqnarray}
in which $\rho$ is the density matrix of the system, $c$ is the dissipation rate, $V_{l\rightarrow m}\equiv|m\rangle \langle l|$ is the jump matrix from the $m$th energy level to the $l$th one, and $W_{l\rightarrow m}$ is the transition probability satisfying the detailed balance condition, $W_{l\rightarrow m}/W_{m\rightarrow l}={\rm exp}[-(E_m-E_l)/T]$. This condition ensures that thermal state will be reached in the long time limit, independent of the detailed form of $W_{l\rightarrow m}$. The dynamics described by the Eq.~(\ref{Lind}) includes contributions from both the quantum and classical thermal fluctuations. The first part in the right hand side of Eq.~(\ref{Lind}) shows the quantum unitary evolution; while the second part of the right hand side gives the master equation describing the classical stochastic process~\cite{Cardy,Tauber}.

For the driven dynamics in the overlapping critical region, we consider the case in which the temperature $T$ is fixed at a classical critical temperature, i.e., $T=T_c$ ($g$ is also fixed according to Eq.~(\ref{Eq:tcg})), and the symmetry-breaking field $h$ is changed linearly. According to the HKZS, the driven dynamics of the order parameter $M$ satisfies the classical KZS~\cite{Zhong1,Zhong2},
\begin{equation}
M(h,R)=R^{\frac{\beta_{\rm cl}}{\nu_{\rm cl} r_{\rm cl}}}f_{9}(hR^{-\frac{\beta_{\rm cl}\delta_{\rm cl}}{\nu_{\rm cl} r_{\rm cl}}}),
\label{Eq:clakzs}
\end{equation}
in which the subscript $\rm cl$ indicates that the critical exponents are the classical ones. For the $2$D classical case, $\beta_{\rm cl}$, $\nu_{\rm cl}$ and $\delta_{\rm cl}$ equals the corresponding ones for the $(1+1)$D FPPT (See Table \ref{tabexp}), while $z_{\rm cl}=2.1667$ and $r_{\rm cl}=4.0417$.

Then the HKZS also shows that the driven dynamics of $M$ also satisfies the $(2+1)$D quantum KZS~\cite{YinZhong},
\begin{eqnarray}
\begin{split}
&M(h,g,T,c,R)\\&=R^{\frac{\beta_{\rm q}}{\nu_{\rm q} r_{\rm q}}}f_{10}(hR^{-\frac{\beta_{\rm q}\delta_{\rm q}}{\nu_{\rm q} r_{\rm q}}},gR^{-\frac{1}{\nu_{\rm q} r_{\rm q}}},TR^{-\frac{z_{\rm q}}{r_{\rm q}}},cR^{-\frac{z_{\rm q}}{r_{\rm q}}}),
\label{Eq:quakzs}
\end{split}
\end{eqnarray}
in which the subscript $\rm q$ indicates that the critical exponents are those for the quantum $(2+1)$D FPPT, and $\beta_{\rm q}\simeq0.326$, $\nu_{\rm q}\simeq0.630$, $\delta_{\rm q}\simeq4.790$ and $z_{\rm q}=1$.

By the similar procedure in Sec.~\ref{sec:HKZS}, we obtain the HKZS constraint on Eq.~(\ref{Eq:quakzs}),
\begin{eqnarray}
\begin{split}
f_{10}&=(T_cR^{-\frac{z_{\rm q}}{r_{\rm q}}})^{-\frac{r_{\rm q}}{z_{\rm q}}(\frac{\beta_{\rm cl}}{\nu_{\rm cl} r_{\rm cl}}-\frac{\beta_{\rm q}}{\nu_{\rm q} r_{\rm q}})}\\&
\times f_{11}[(hR^{-\frac{\beta_{\rm q}\delta_{\rm q}}{\nu_{\rm q} r_{\rm q}}})(T_cR^{-\frac{z_{\rm q}}{r_{\rm q}}})^{-\frac{r_{\rm q}}{z_{\rm q}}(\frac{\beta_{\rm q} \delta_{\rm q}}{\nu_{\rm q} r_{\rm q}}-\frac{\beta_{\rm cl} \delta_{\rm cl}}{\nu_{\rm cl} r_{\rm cl}})}],
\label{Eq:quakzs1}
\end{split}
\end{eqnarray}
for fixed $T_c/c$.

However, we note that the Lindblad equation in Eq.~(\ref{Lind}) contains all the eigenstates in principle and may thus be difficult to be solved even in $(1+1)$D~\cite{Attal,YinZhong}. Although the $2D$ quantum Ising model at finite temperatures is a very common model in condensed matter physics, the numerical study on its critical dynamics for large lattice sizes is hindered by the lack of effective numerical methods~\cite{Cirac}. Nonetheless, since real experiments is implemented at finite temperatures, our scaling theory of the HKZS in $2$D quantum Ising model may be examined experimentally.

\subsection{\label{discu2}A generalization of the HKZS}
We can also consider more complex cases in which the overlapping region is constituted by more than two critical regions. For example, we can also consider the participation of the critical region of the $(1+1)$D YLES~\cite{YinChen}. So, there is an overlapping region involving the $(0+1)$D YLES, $(1+1)$D YLES and the $(1+1)$D FPPT. According to the HKZS, the driven dynamics should be described by all three critical theories. In principle, one can also obtain the constraint on the scaling functions in this complex overlapping critical region. However, in real situation, to realize such a complex overlapping region, one should carefully tune more parameters. We leave this for further studies.

\section{\label{sum}Summary}
In summary, we have studied the driven dynamics in the overlapping critical regions. We have suggested a HKZS to describe the scaling behavior in the driven process. By assuming that the overlapping region is constituted by the critical Region-A and critical Region-B, we have shown that, according to the HKZS, the driven dynamics in the overlapping region can be described by the critical theories for both Region-A and Region-B simultaneously. This results in a constraint on the scaling function in the overlapping critical region. We have verified the HKZS by numerically solving the driven dynamics in the quantum Ising chain with an imaginary longitudinal field. We have also discussed the HKZS in the $2$D quantum Ising model, in which the HKZS may be examined experimentally. Although the Hamiltonian of model~(\ref{model}) is non-Hermitian, the static scaling of the YLES has been detected in various experiments~\cite{Binek,WeiLiu,PengLiu}. Maybe the HKZS can be detected in these systems by adjusting the system to the vicinity of its critical point and imposing an external time-dependent field.

\section*{Acknowledgments}
This work is supported by the National Natural Science Foundation of China (grant numbers 11704161 and 11547142) and the Natural Science Foundation of Jiangsu Province of China (Grant No. BK20170309).

\end{CJK*}

\end{document}